\documentclass[numreferences]{preprint}
\usepackage{epsfig,preprintcite}

\begin{document}
\begin{article}

{\ }
\vspace{10cm}
\begin{center}
This is an unedited preprint. The original publication is available at \\
http://www.springerlink.com \\
http://www.doi.org/10.1007/s10751-006-9242-4
\end{center}

\begin{opening}
\title{Dynamics of Metal Centers \\ Monitored by Nuclear Inelastic Scattering}            

\author{H. \surname{Paulsen}}
\author{P. \surname{Wegner}} 
\author{H. \surname{Winkler}} 
\author{J. A. \surname{Wolny}} 
\author{L. H. \surname{B{\"o}ttger}} 
\author{A. X. \surname{Trautwein}} 
\institute{Institut f{\"ur} Physik, Universit{\"a}t zu L{\"u}beck, D-23538 L{\"u}beck, Germany
           \email{paulsen@physik.uni-luebeck.de}}                               
\author{C. \surname{Schmidt}}
\institute{Institut f{\"ur} Biochemie, Universit{\"a}t zu L{\"u}beck, D-23538 L{\"u}beck, Germany}
\author{V. \surname{Sch{\"u}nemann}}
\institute{Fachbereich Physik, Technische Universit{\"a}t Kaiserslautern, D-67663 Kaiserslautern, Germany}
\author{G. \surname{Barone}}
\author{A. \surname{Silvestri}}
\institute{Dipartimento di Chimica Inorganica e Analitica, Universit\'a degli Studi di Palermo, I-90128 Palermo, Italy}
\author{G. \surname{La Manna}}
\institute{Dipartimento di Chimica Fisica, Universit\'a degli Studi di Palermo, I-90128 Palermo, Italy}
\author{A.I. \surname{Chumakov}}
\author{I. \surname{Sergueev}}
\author{R. \surname{R{\"u}ffer}}
\institute{ESRF, F-38043 Grenoble Cedex 9, France}


\runningtitle{Dynamics of Metal Centers Monitored by Nuclear Inelastic Scattering}
\runningauthor{Paulsen et al.}

\begin{abstract} 
Nuclear inelastic scattering of synchrotron radiation has been used now since ten years
as a tool for vibrational spectroscopy.
This method has turned out especially useful in case of large molecules that contain
a M{\"o}ssbauer active metal center.
Recent applications to iron-sulfur proteins, to iron(II) spin crossover complexes and to
tin-DNA complexes are discussed.
Special emphasis is given to the combination of nuclear inelastic scattering and density
functional calculations.
\end{abstract}


\abbreviations{\abbrev{NIS}{nuclear inelastic scattering}; \abbrev{SCO}{spin crossover};
               \abbrev{DFT}{density functional theory}; \abbrev{EXAFS}{extended X-ray absorption fine structure};
               \abbrev{msd}{mean square displacement}; \abbrev{HS}{high-spin}; \abbrev{LS}{low-spin}}

\end{opening}

\section{Introduction}
Nuclear inelastic scattering (NIS) of synchrotron radiation (also known as nuclear resonant
vibrational spectroscopy, NRVS) can be regarded as an extension of the conventional, energy-resolved
M{\"o}ssbauer spectroscopy to energies of the order of molecular vibrations
\cite{Seto95,Stuhrhahn95}.
NIS has similarity with resonance Raman spectroscopy: the energy of monochromatized
synchrotron radiation is detuned from nuclear resonance by a certain amount of energy (several meV)
that is sufficient to create or annihilate a phonon, or in terms of molecular spectroscopy, to excite
or deexcite a molecular vibration.
While the Raman intensity for a given molecular vibration depends on the change of polarizability with respect
to the motion of the nuclei, the NIS intensity depends on the contribution of this particular vibration to the
mean-square displacement (msd) of the M{\"o}ssbauer nucleus.
NIS is, therefore, a highly site-selective tool and it is as vibrational spectroscopy complementary to Raman
or IR spectroscopy.
The interpretation of measured NIS spectra is greatly facilitated due to the fact that the NIS intensity
depends only on mechanical and not on electric properties of the investigated molecules.
For example, iron-ligand bond stretching vibrations of an iron-containing molecule are normally easy to identify
because of their relatively large contribution to the msd.
The dependence on purely mechanical properties also facilitates the simulation of measured NIS spectra with the help
of quantum chemical calculations, which are usually based on density functional theory (DFT).
The calculation of IR or Raman intensities, which depend on  electric properties like changes of dipole moment or
polarizability tensor, is far less accurate.
A quantitative analysis of this fact with the example of the ferrocene complex is given in section \ref{DFT}.

The comparison of experimental and simulated spectra provides the possibility to assign peaks in
the measured NIS spectrum with calculated normal modes of the molecule under study.
This procedure has been applied first to small and medium-sized metal complexes (up to 100 atoms)
\cite{Paulsen99,Paulsen01} but is now applied also to larger biological molecules like rubredoxin
\cite{Trautwein05} or DNA with chelated metal ions \cite{Barone05} (in the latter two cases DFT calculations
are restricted to a cluster of about 100 atoms around the iron center).
First applications of NIS to problems of biological relevance were related to model hemes like
tetraphenylporphyrins \cite{Sage01,Leu04} and to heme-based proteins like myoglobin
\cite{Sage01,Leu04,Achterhold96} (see also references in \cite{Paulsen05}).
Recently NIS measurements have been performed on model complexes of iron-sulfur proteins
\cite{Oganesyan04}, on a rubredoxin-type iron-sulfur protein in the reduced and in the oxidized
state \cite{Trautwein05}, and on Benzoyl-CoA reductase \cite{Trautwein05},
which contains [4Fe-4S]$^{2+}$ centers.
The NIS spectra for the rubredoxin-type protein allowed for the first time to investigate the
change of the metal dynamics between the reduced and the oxidized form
(see also sections \ref{FeS} and \ref{TinDNA}).

Another class of metal-containing molecules, which is being studied by NIS, is formed by the so called
spin crossover (SCO) complexes.
For these molecules NIS has turned out to be a very valuable method, since the dynamics of the metal center
governs the crossover from the low-spin to the high-spin isomer (section \ref{SCO}).

\section{Density functional theory calculations} \label{DFT}
A detailed analysis of the dynamics of metal centers in large molecules requires
combination of experimental investigations and theoretical calculations.
In case of molecules that contain transition metals, density functional theory (DFT) calculations
are currently the method of choice.
They have been shown to reproduce the frequencies and vibrational modes of transition metal compounds
with reasonable accuracy and can be applied also to large molecules containing hundreds of atoms.
Frequencies and vibrational modes are used to simulate a NIS spectrum for comparison with experiment
\cite{Paulsen99,Sage01}.
Additionally, DFT calculations provide the partial vibrational density of states (PVDOS)
for the M{\"o}ssbauer nucleus which can be compared with a PVDOS that has been extracted from the measured spectrum.
A detailed description how the calculated vibrational modes can be used to simulate NIS spectra can be found
for instance in Refs. \cite{Paulsen99,Sage01}.

In a low-temperature approximation the absorption probability $S$ that is measured in a NIS experiment
is proportional to the spectral contribution to the mean-square displacement (msd) of the M{\"o}ssbauer
nucleus:
\begin{equation}
   S(E,{\bf k}) \propto \sum_i \delta(E-h\nu_i) \left\langle \left({\bf k} \cdot {\bf u}_i \right)^2 \right\rangle
\end{equation}
Here $E$ and $\bf k$ denote energy and wavevector, respectively, of the absorbed photon, index $i$ enumerates
calculated normal modes of vibrations.
$\nu_i$ is the frequency of mode $i$ and the vector ${\bf u}_i$ contains those three components of the normal
coordinate vector that refer to the motion of the M{\"o}ssbauer nucleus in $x$, $y$ and $z$ direction.
The sum over all terms $\left\langle \left({\bf k} \cdot {\bf u}_i \right)^2 \right\rangle$ yields the
total measured msd of the M{\"o}ssbauer nucleus.
In order to simulate the NIS spectrum of a molecule it is, therefore, sufficient to know the frequencies $\nu_i$
and to which degree the M{\"o}ssbauer nucleus is participating in the individual vibrations.
The latter information is given by the vectors ${\bf u}_i$.
The frequencies and normal modes of molecular vibrations may be regarded as mechanical properties of the molecule,
although they are based on the molecule's electronic structure.

\begin{figure}[t]
      \centerline{\includegraphics[width=1.1\textwidth]{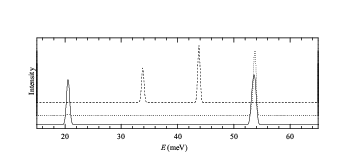}}
      \caption{NIS, IR and Raman intensities (solid, dotted and dashed line, respectively) of ferrocene in arbitrary units
               calculated with B3LYP/TZVP.}
      \label{tzvp}
\end{figure}

\begin{figure}[t]
      \centerline{\includegraphics[width=1.1\textwidth]{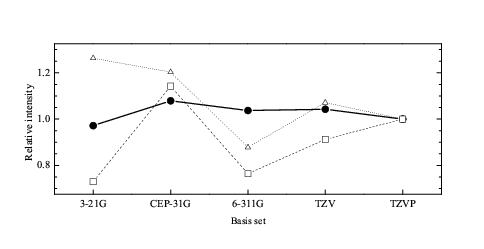}}
      \caption{Relative NIS ($\bullet$), IR ($\Box$) and Raman ($\triangle$) intensities of mode 4 (Raman) and
               the nearly degenerate modes 8 and 9 (IR, NIS) of ferrocene (see text) calculated with B3LYP and
               different basis sets as indicated.
               For each spectroscopic technique the calculated intensities are divided by the value obtained with
               the TZVP basis set.
               The lines serve as guides to the eyes.}
      \label{basisset}
\end{figure}

With the example of the ferrocene molecule we demonstrate here that these mechanical properties are more easily
calculated with sufficient accuracy than those electric properties (changes of dipole moment or polarizability
tensor) that are needed for the simulation of IR or Raman spectra.
The dynamics of the iron center in ferrocene has been studied intensively by various spectroscopic techniques
including NIS \cite{Jayasooriya01} and quasielastic nuclear forward scattering \cite{Asthalter03}.
We have performed DFT calculations using the hybrid functional B3LYP \cite{Becke93} together with five different
basis sets, 3-21G \cite{Binkley80,Gordon82,Pietro82,Dobbs86,Dobbs87,Dobbs87a},
CEP-31G \cite{Stevens84,Stevens92,Cundari93}, 6-311G \cite{McLean80,Krishnan80} (with the Wachters-Hay double zeta
basis for Fe \cite{Wachters70,Hay77}), TZV and TZVP \cite{Schaefer92,Schaefer94} as included in the Gaussian 03
program package \cite{Gaussian03}.
The simulated spectra (Fig.~\ref{tzvp}) are in reasonable agreement with the experimental spectra given in ref.
\cite{Jayasooriya01}.
The comparison of calculated IR, Raman and NIS spectra in the region of metal-ligand vibrations exemplifies the
complementarity of NIS and Raman spectra.
In the region from 0 to 60~meV nine vibrational modes are obtained from the calculation.
Modes 1 and 7 are practically invisible for all three spectroscopic techniques, modes 2 and 3, modes 5 and 6 and
modes 8 and 9 are in pairs nearly degenerate.
In this region mode 4 has the strongest Raman intensity, while the almost degenerate pair of modes 8 and 9
exhibits the largest IR and NIS intensities.
We have examined the influence of the basis set on the calculated IR, Raman and NIS intensities.
In Fig.~\ref{basisset} the basis sets are ordered on the horizontal axis with increasing quality.
It turns out that the calculated NIS intensity is almost independent of the employed basis set whereas the
calculated IR and Raman intensities exhibit large oscillations for small and medium sized basis sets.
This result is even more striking if one takes into account that, according to Lambert-Beer law, there is an
exponential relation between the calculated IR and Raman activities and the measured spectral intensities.
In summary, a reasonable theoretical simulation of IR and, especially, of Raman spectra is computational much
more demanding than a simulation of NIS spectra.
This is an important aspect if large molecules are investigated where demanding computational methods are
prohibitive.

\section{Recent examples}

\subsection{Rubredoxin-type iron-sulfur proteins} \label{FeS}
Iron-sulfur proteins are widely used for electron transport in cells.
For a deeper understanding of the reorganisation of these proteins upon electron uptake or release
the knowledge of the vibrational modes of the iron-sulfur core is of special importance.
For the oxidized and reduced forms of a rubredoxin-type iron sulfur protein from {\sl Pyrococcus abyssi},
which contains a single Fe-S$_4$ core \cite{Czernuszewicz86}, we have measured NIS spectra at
the nuclear resonance beamline ID 18 of ESRF in Grenoble, France.
In the NIS spectrum of the oxidized form Fe-S bond-stretching modes could be identified which are in agreement
with resonance Raman studies.
The spectrum of the reduced form, where no Raman data are available, shows a down shift of these bond-stretching
modes by about 5 meV.
This observation is in line with an increase of the Fe-S bond length upon reduction as observed by
EXAFS.
In addition to the peaks that are assigned to the Fe-S bond stretching modes, a second group of prominent peaks
was observed in the NIS spectra, which hardly change their positon upon reduction.
In order to assign these peaks to molecular vibrations, DFT calculations have been performed for a series of models
of the protein under study, Fe(SH)$_4$, Fe(SCH$_3$)$_4$ and FeCys$_4$, i.e. containing 9, 21 and 49 atoms, respectively.
The simulated NIS spectra for these three models gain approach to the experimental spectra (oxidized and reduced) with
growing model size and allow to assign the observed second group of peaks to Fe-S-C bending modes.
The most simple model, Fe(SH)$_4$, although far less realistic than the larger models, is nevertheless useful
since it provides a qualitatively correct picture of the experimentally observed vibrations
of the FeS$_4$ core.
More details of this study are described in ref. \cite{Trautwein05}.

\subsection{Iron corroles}
Corroles are aromatic macrocycles that are closely related to porphyrins and have unique chemical properties such
as stabilizing high oxidation states of coordinated metal ions, e.g. Fe(IV) \cite{Zakharieva02}.
We have performed NIS measurements on iron corroles with two different axial ligands at
the nuclear resonance beamline ID 18 of ESRF.
Currently, a detailed analysis of the observed molecular vibrations of these molecules is performed that exploits the
complementarity of IR, Raman and NIS measurements in combination with DFT calculations \cite{Tuczek05}.

\subsection{Tin-DNA complexes} \label{TinDNA}
NIS is a highly site-selective tool, focussed on M{\"o}ssbauer nuclides. Therefore, most applications of NIS in
biology and bioinorganic chemistry have so far been performed with molecules containing iron centers.
Recently, we have investigated tin-DNA complexes at the nuclear resonance beamline ID 22 of ESRF;
this is, to our knowledge, the first example of a NIS study on a biological sample that contains
a M{\"o}ssbauer nuclide other than $^{57}$Fe.
This study has been performed to elucidate the binding site of organotin cations, Sn(II) and Sn(IV), chelated to DNA.
The binding site of such tin-DNA complexes could not be determined by conventional structural methods like X-ray
diffraction.
Spectroscopic techniques like IR or Raman are hampered by the complexity of the spectra which are dominated by the
vibrations of the large organic ligands.
The site-selectivity of NIS focusses on those vibrations that are connected with a considerable msd
of the tin nucleus.
Our NIS investigation in combination with DFT calculations have corroborated the former hypothesis
about the binding sites of tin that was based on conventional M{\"o}ssbauer studies and semiempirical
molecular-orbital calculations \cite{Barone99}.
More details can be found in refs. \cite{Barone05,Barone05a}.

\subsection{Spin crossover complexes} \label{SCO}
These molecules exhibit a transition from a low-spin (LS) to a high-spin (HS) isomer that is induced by
change of temperature or pressure or by irradiation with light \cite{Guetlich04}.
Spin crossover complexes are not only very promising materials for technical applications,
they also serve as models for the transition of the metal spin in biomolecules,
observed for instance in a catalytic cycle.
The driving force for the spin transition is the entropy difference between the two spin isomers.
The major contribution to this entropy difference arises from the dynamics of the metal ion: in the
LS state the $d$ electrons of the metal center occupy non-bonding or weakly $\pi$-anti-bonding molecular
orbitals.
Upon transition to the HS isomer part of the $d$ electrons are transferred to $\sigma$-anti-bonding molecular
orbitals.
Consequently, in the HS isomer the metal-ligand bonds are weaker and longer and also softer, i.e. the
force constants of the metal-ligand bond stretching vibrations are lower.
This leads to a significant down-shift of the corresponding frequencies upon crossover from LS to the HS state
(about 10 to 20~meV in case of typical Fe(II) SCO complexes).
For this reason, the HS isomer has - in comparison to the LS isomer - many more vibrational states
that are thermally accessible, and the vibrational entropy contribution of the HS isomer is significantly
higher than the corresponding value of the LS isomer (some 10~J~mol$^{-1}$~K$^{-1}$ in case of Fe(II) SCO
complexes).
First NIS experiments have been performed on samples of [Fe(tpa)(NCS)$_2$] (tpa = tris(2-pyridylmethyl)amine)
and [Fe(bpp)$_2$](BF$_4$)$_2$ (bpp = 2,6-bis(pyrazol-3-yl)pyridine) \cite{Paulsen99,Paulsen01a,Chumakov00}.
Subsequent measurements on a single crystalline sample of [Fe(tptMetame)](ClO$_4$)$_2$
(tptMetame = 1,1,1-tris((N-(2-pyridylmethyl)-N-methylamino)methyl)ethane) yielded angular resolved NIS spectra
\cite{Paulsen01}.
With the same sample the influence of a pressure induced SCO on the metal dynamics has been investigated with
NIS \cite{Averseng05}.
Studying [Fe(phen)$_2$(NCS)$_2$] (phen = 1,10-Phenanthroline) the complementarity of NIS and IR and
Raman spectroscopy has been demonstrated.
Combining these three vibrational spectroscopic techniques with DFT calculations allows an almost complete
assignment of the observed vibrational modes \cite{Ronayne05}.
One of the most recent NIS experiments on SCO complexes \cite{Boettger05} has been used to investigate five
different phases of the complex [Fe(ptz)$_6$](BF$_4$)$_2$ (ptz = 1-$n$-Propyl-tetrazole) that can be reached
by combination of temperature changes and irradiation with light of different frequencies \cite{Kusz04}.

\section{Conclusion}
In summary we conclude that the comparison of complementary experimental (i.e. NIS, IR, Raman) and theoretical
spectra provides the possibility to assign measured peaks with calculated normal modes of the metal complex
under study.
This procedure has been applied to small and medium-sized iron complexes and has recently been applied also
to iron-containing proteins and to DNA being chelated with tin.

\begin{acknowledgements}
Financial support by the Deutsche Forschungsgemeinschaft (DFG) is gratefully acknowledged
(Tr 97/31 and Tr 97/32). 
\end{acknowledgements}

\end{article}
\end{document}